\documentstyle[preprint,floats,tighten,epsf,prd,aps]{revtex}

\def\be{\begin{equation}}
\def\ee{\end{equation}}

\def\bea{\begin{eqnarray}}
\def\eea{\end{eqnarray}}
\def\bml{\begin{mathletters}}
\def\blea{\begin{mathletters}\begin{eqnarray}}
\def\elea{\end{eqnarray}\end{mathletters}}

\begin{document}
\draft
\title{Monopole-antimonopole bound states as a source of
ultra-high-energy cosmic rays.}

\author{J.\ J.\ Blanco-Pillado\footnote{Email address: {\tt
jose@cosmos2.phy.tufts.edu}} and Ken D. Olum\footnote{Email address: {\tt kdo@alum.mit.edu}}}

\address{Institute of Cosmology \\
Department of Physics and Astronomy \\
Tufts University \\
Medford, MA 02155}

\date{April 1999}

\maketitle

\begin{abstract}%
The electromagnetic decay and final annihilation of magnetic
monopole-antimonopole pairs formed in the early universe has been
proposed as a possible mechanism to produce the highest energy cosmic
rays.  We show that for a monopole abundance saturating the Parker
limit, the density of magnetic monopolonium formed is many orders of
magnitude less than that required to explain the observed cosmic ray
flux.  We then propose a different scenario in which the monopoles and
antimonopoles are connected by strings formed at a low energy phase
transition ($\sim 100 \, \text{GeV}$).  The bound states decay by gravitational
radiation, with lifetimes comparable with the age of the universe.
This mechanism avoids the problems of the standard monopolonium
scenario, since the binding of monopoles and antimonopoles is
perfectly efficient.
 
\end{abstract}


\newpage

\section{Introduction}
The observation of ultra-high-energy cosmic rays (UHECR) with energies
above $10^{11} \, \text{GeV}$\cite{Haya94,Bird94} poses a serious challenge to
the particle acceleration mechanisms so far proposed.  This fact has
motivated the search for non-acceleration models, in which the high
energy cosmic rays are produced by the decay of a very heavy
particle. Topological defects are attractive candidates for this
scenario.  Due to their topological stability these objects can retain
their energy for very long times and release quanta of their
constituents, typically with GUT scale masses, which in turn decay to
produce the UHECR.

Various topological defect models and mechanisms have been studied by
numerous authors\cite{Pijus98}. In this paper we investigate two
different scenarios involving the annihilation of
monopole-antimonopole pairs. We first discuss standard magnetic
monopole pair annihilation \cite{Hill83,Bhatta95}, paying particular
attention to the kinetics of monopolonium formation.  We find that, due
to the inefficiency of the pairing process, the density of
monopolonium states formed is many orders of magnitude less than the value
required to explain the UHECR events.
 
We then present a different scenario in which very massive monopoles
($ m \sim 10^{14} \, \text{GeV}$) are bound by a light string formed at
approximately $100\, \text{GeV}$.  These monopoles do not have the usual
magnetic charge, or in fact any unconfined flux.  Gravitational
radiation is the only significant energy-loss mechanism for the
bound systems.\footnote{Such systems were studied in a different context 
by Martin and Vilenkin\cite{Martin:1997cp}.}
Their lifetimes can then be comparable with the age
of the universe, and their final annihilation will then contribute 
to the high energy end of the cosmic ray spectrum.

\section{Required monopolonium abundance.}

What density of decaying monopolonium states is required to produce the
observed cosmic rays?  The monopolonium will behave as a cold dark
matter (CDM) component and will cluster in the galactic halo,
producing a high energy spectrum of cosmic rays without the
Greisen-Zatsepin-Kuzmin (GZK) cutoff\cite{Greisen,Zatsepin}.  Since
the observational data does not seem to show any such cutoff, this is
an advantage of such topological defect models \cite{Bere97,Bere98}.

For a given monopole mass, we can set the lifetime of the monopolonium
at least equal to the age of the universe, and obtain the required
density of monopolonium in the halo by normalizing the flux to the
observed high energy spectrum\cite{Bere97}.  The required number
density decreases with the monopole mass, so as a lower limit we can
take the required density corresponding to 
$m_M =10^{17}\, \text{GeV}$\cite{Bere97},
\be
N^{h} _{M \bar M} (T_0) > 6\times 10^{-27} \, \text{cm}^{-3}\,.
\ee

Since the different components of the CDM cluster in the same way we
can use this halo density to get the mean density in the universe, 
by computing,

\be
N_{M \bar M} = {{N^h_{M \bar M} \, \Omega_{CDM} \, 
\rho_{cr}}\over {\rho^{h}_{CDM}}}.
\ee

For $\Omega_{CDM} h^2 = 0.2$, $\rho^{h}_{CDM}= 0.3\, \text{GeV}$
$\text{cm}^{-3}$, and $\rho_{\text{cr}} = 10^4 h^2\,\text{eV cm}^{-3}$,
we get
\be
N_{M \bar M} (T_0) > 10^{-32} \,  \text{cm}^{-3}\,.
\ee

\label{sec:needed}
We will work with a comoving monopolonium density
 $\Gamma = N_{M\bar M}/s$ where $s$ is the entropy density, currently 
$s \approx 3 \times 10^{3} \,  \text{cm}^{-3}$, so that we require
\be\label{eqn:Gammaneeded}
\Gamma > 10^{-35}
\ee
to explain the observed UHECR.


\section{Magnetic monopole states}
\subsection{Introduction}
Monopolonium states are expected to have been formed by radiative
capture if there was a non-zero density of free monopoles in the early
universe.  They will typically be bound in an orbit with a large
quantum number, so we can treat them as classical objects emitting
electromagnetic radiation as they spiral down to deeper and deeper
orbits, until they annihilate in a final burst of very high energy
particles.

The electromagnetic decay of monopolonium was analyzed by
Hill\cite{Hill83} using the dipole radiation formula. The rate of
energy loss is\footnote{Here and throughout we use units where $\hbar
= c = k_B = 1$.}
\be
{{dE}\over{dt}} = {{64 \, E^4}\over{3 \, g_M^2 m_M^2}}\,,
\ee
where $g_M$ is the magnetic charge.
From this expression, the lifetime of monopolonium with radius $r$ and
binding energy $E = g_M^2/2r$ is\cite{Hill83}
\be
\tau_E \sim {m_M^2 \, r^3\over {8 g_M^4}}\,.
\ee
For $m_M = 10^{16}\, \text{GeV}$, $g_M = 1/(2e) \approx\sqrt {34}$, and
an initial radius of $r= 10^{-9} \, \text{cm}$, this gives 
$\tau_E\sim 10^{18} \, \text{sec}$, comparable to the age of the universe.

Bhattacharjee and Sigl\cite{Bhatta95} used a thermodynamic equilibrium
approximation to estimate the monopolonium density and argued that the
late annihilation of very massive magnetic monopoles could explain the
UHECR events observed.  Here we recalculate the density of monopolonium
states, taking into account the kinematics of formation and the 
frictional energy loss of monopolonium formed at early times.

\subsection{Friction}
Before electron-positron annihilation, monopoles interact with a 
background of relativistic charged particles.  These interactions produce a 
force which, for a non-relativistic monopole is given by \cite{Kolb81}
\be\label{eqn:F}
F = {\pi\over 18}N_c  T^2 v 
\int^{b_{\text{max}}}_{b_{\text{min}}} {db\over b}
\ee
where $N_c$ is the number of species of charged particles, $v$ the
velocity of the monopole with respect to the background gas of charged
particles and $b$ the impact parameter of the incident particles.
Since we are interested in the friction that a monopole feels in a
bound state orbit of monopolonium, we will not consider the
interaction of charged particles with impact parameter greater than
the radius of the monopolonium, so $b_{\text{max}} \approx g_M^2
E^{-1}$.  Initially, the monopoles are bound with energy $E\sim T$, so
$b_{\text{max}}\approx g_M^2T^{-1}$.  Equation (\ref{eqn:F}) is
derived using the approximation that each charged particle is only
slightly deflected.  This approximation breaks down for impact
parameters that are too small, so we should cut off our integration at
\cite{Kolb81} $b_{min} \approx T^{-1}$.  Using $N_c = 2$ and $g_M^2
\approx 34$, we get
\be
F \approx 1.22 \, T^2 v
\ee
so the energy loss rate due to interactions with charged particles in
the background is
\be
{{dE}\over{dt}} \approx 1.22 \, T^2 v^2.
\ee

Taking the system to be bound in a circular orbit, we have
\be
m_M v^2 \sim E
\label{eqn:e}
\ee
so we can write
\be
{{dE}\over{dt}} \approx   1.22 \, T^2 \, {E\over {m_M}}
\label{eqn:dedtf1}
\ee
The time scale for this process is
\be
\tau_F={E\over dE/dt} \approx {{m_M}\over {1.22 \, T^2}}
\ee
If we compare it with the Hubble time,
\be
\tau_H =\sqrt {90\over 8\pi^3 g_*} m_{pl} T^{-2}\approx 0.184 m_{pl} T^{-2}\,,
\label{eqn:time}
\ee
where $m_{pl}$ is the Planck mass, and $g_*$ is the number of effectively 
massless degrees of freedom, $g_* = 10.75$, we get
\be
{{\tau_F}\over {\tau_H}} \approx 0.15 \, {{m_M}\over {m_{pl}}} \ll 1.
\ee
Thus, we see that the damping of the monopolonium energy due to
friction is very effective in this regime, and the monopoles spiral
down very quickly.  

When the distance between monopoles becomes small
as compared to $T^{-1}$, the effect of friction is reduced and Eq.\
(\ref{eqn:F}) is no longer accurate.  However, even for $T =
1 \, \text{MeV}$, the radius has been reduced about two orders of
magnitude to $r\sim 2\times 10^{-11}\, \text{cm}$, and the
electromagnetic lifetime has been reduced by about six orders of
magnitude.  Thus only monopolonium states formed after electron-positive
annihilation can live to decay in the present era.

After electron-positron annihilation the number of charged particles in the 
thermal background has decreased by a factor $\sim 10^{-9}$ so
$\tau_F/\tau_H\gg 1$ and the monopolonium is little affected by
friction.

\subsection{Formation rate}
We can obtain an upper limit for the monopolonium density by solving
the Boltzmann equation,
\be
{dN_{M \bar M}\over{dt}} =  \langle \sigma_b v \rangle n_{M}^2 - 
3 \, H\,  N_{M \bar M},
\ee
where $n_M$ denotes the free monopole density, $N_{M \bar M}$ 
the monopolonium density, $H$ the Hubble constant, and 
$\langle \sigma_b v \rangle$ the average product of the binding 
cross section times the thermal velocity of the monopoles. 

With the comoving monopole density $\gamma = n_{M}/s$,
we can rewrite the equation above as
\be
{{d \Gamma}\over{dt}} =    \langle \sigma_b v \rangle {\gamma} \,
n_{M} = \langle \sigma_b v \rangle \gamma^2 s \,. 
\ee
Using the approximation for the classical radiative capture cross 
section of monopoles with thermal velocities given by \cite{Dicus82},
\be
\langle \sigma_b v \rangle \approx {{\pi^{7/5}}\over 2} \,  {{{g_{M}}^4}\over {{m_M}^2}}
 {\left({{m_M}\over T}\right)}^{9/10}\,,
\label{eqn:radiative}
\ee
and with
\be s ={2\pi^2\over 45} g_{*S} T^3\,,
\ee
where $g_{*S}$ is the number of degrees of freedom contributing to the
entropy, we get
\be
{{d \Gamma}\over{dt}} = {{\pi^{17/5}}\over 45} 
  {g_{M}^4 \gamma^2 \over m_M^2}
{\left({{m_M}\over T}\right)}^{9/10}
g_{*S} T^3\,.
\ee
Since we are interested in the evolution of the monopolonium density
after electron-positron annihilation, we will take a constant value
$g_{*S}\approx 3.91$ to get
\be
{{d \Gamma}\over{dt}} \approx 4.25 {g_{M}^4 \gamma^2 \over m_M^2}
{\left({{m_M}\over T}\right)}^{9/10} T^3\,.
\label{eqn:dgdt}
\ee

As we will see, only a tiny fraction of the monopoles will ever be
bound, so we can consider the comoving number of monopoles $\gamma$ to
be constant.  To integrate Eq (\ref{eqn:dgdt}), we will make a change
of variable
\be
t =\sqrt {90\over 32\pi^3 g_*} m_{pl} T^{-2}\approx 0.164m_{pl} T^{-2}\,,
\ee
appropriate to times after electron-positron annihilation,
to get
\be
{{d \Gamma}\over{dT}} \approx -1.34  {g_{M}^4 m_{pl} \gamma^2 \over m_M^2}
{\left({{m_M}\over T}\right)}^{9/10}\,.
\ee
and thus
\be
\Gamma_f \approx 13.4 g_M^4 \left({{m_{pl}}\over {m_M}}\right)
  {\left({{T_i}\over {m_M}}\right)}^{1/10} {\gamma}^2\,.
\ee
We now take $T_i \sim 1 \, \text{MeV}$ and $g_M^2=34$, and note 
that to produce the observed UHECR, we must have 
$m_M > 10^{11}\, \text{GeV}$, so that for a fixed 
monopole comoving density $\gamma$, we have the bound,
\be
\Gamma_f < 4\times 10^6 {\gamma}^2\,.
\label{eqn:Gammaf}
\ee
\subsection{Monopole density bound}
The formation of magnetic monopoles via the Kibble
mechanism\cite{Kibble76} is inevitable in all GUT models of the early
universe, and annihilation mechanisms are not efficient in a
rapidly expanding background\cite{Zeldovich78,Preskill79}, so that the
typical initial density of monopoles produced at a GUT phase
transition will very soon dominate the energy density of the universe.
The most attractive solution for this problem is the inflationary
scenario\cite{Guth81}. In standard inflation, the exponential expansion 
of the universe reduces the monopole density to a completely negligible value.
 However, it is possible for new monopoles to be formed at the end of
inflation\cite{Turner82,Shafi:1984tt,Shafi:1984bd,Kuzmin99,Tkachev:1998dc}.
The exact relic abundance of monopoles created in this period 
is very model dependent, but its value is constrained by the
 Parker limit \cite{Parker70}: To prevent the acceleration of 
monopoles from eliminating the galactic magnetic field, the monopole 
flux into the galaxy must be limited by
\be
F < 10^{-16}\, \text{cm}^{-2} \, \text{s}^{-1} \, \text{sr}^{-1}\,.
\ee
Assuming a monopole velocity with respect to the galaxy of $ \sim
10^{-3}c$, we can translate this bound into a limit on the monopole
density,
\be
n_{M} < 10^{-23} \, \text{cm}^{-3}\,,
\ee
and thus $\gamma < 10^{-26}$.Then, from Eq.\ (\ref{eqn:Gammaf}) we have
\be
\Gamma_f < 10^{-45}\,.
\ee
Since this conflicts with Eq.\ (\ref{eqn:Gammaneeded}) by 10 orders of
magnitude, we conclude that primordial bound states of magnetic
monopoles cannot explain the UHECR.

We note that we have used several approximations which overstate the
possible value of $\Gamma_f$: First, we have considered the total
classical radiative capture cross section.  This takes into account
not only the monopolonium formed with the right energy to decay at
present, but all the possible binding energies, clearly overestimating
the value of ${\Gamma}_f$.  Second, it has been argued that the
classical cross section given in Eq (\ref{eqn:radiative})
overestimates its real value due to photon discreteness
effects\cite{Dicus82}.  Finally, some of the monopolonium will have
decayed before the present time, reducing the value of $\Gamma$.  All
of these effects make the conflict above more serious.

\section{Monopoles connected by strings.}

We present now a different scenario for the formation and annihilation
of monopole-antimonopole bound states. The main problem in explaining
the UHECR by the conventional magnetic monopolonium system is the
inefficiency of the binding mechanism. This can be solved if we
assume that all the monopoles get connected by strings in a later
phase transition.  Since the U(1) symmetry of the monopoles would be
broken by the second phase transition, this U(1) must be a field other
than the usual electromagnetism.\footnote{This is different from the
Langacker-Pi scenario\cite{Langacker:1980kd}, where electromagnetism 
is broken and then restored at a lower temperature, and monopoles 
do feel large frictional forces.} We furthermore assume that these
monopoles will not have any other unconfined charge, so that they will
feel almost no frictional force moving in a background of
particles.

We take the comoving density of bound monopole systems $\Gamma$ to be
constant.  With a monopole mass of $10^{14}\, \text{GeV}$ the calculation of
Sec.\ \ref{sec:needed} gives $\Gamma\sim 10^{-33}$, and with all
monopoles bound, $\gamma = 2\Gamma$.  The proper density at the time
of string formation is then
\be
n_M (T_s) =\gamma s ={2\pi^2\over 45} g_{*S} T_s^3 \gamma\sim 10^{-32}
T_s^3\,.
\ee
We can then compute the mean separation between monopoles at the time
the string is formed, 
\be
L_i \sim {\left[n_{M}(T_s)\right]}^{-1/3}\,.
\ee
If we take $T\sim 100\, \text{GeV}$, we obtain
\be
L_i \sim 10^{-6} \, \text{cm},
\ee
which is much smaller than the horizon distance, $d_H \sim 3$ cm at
$T\sim 100\, \text{GeV}$.  We will assume that there are no light ($m\sim T_s$
or less) particles that are charged under the string flux.  This means
that there will be no charged particles that interact with the
monopoles and cause the system to lose energy, so that gravitational
radiation will be the only energy loss mechanism.  When the strings
are formed they may have excitations on scales smaller than the
distance between monopoles, but these will be quickly smoothed out by
gravitational radiation, leaving a straight string.  The energy stored
in the string is then $\mu L_i$, where $\mu \sim {T_s}^2$ is the
energy per unit length of the string.  This is smaller than the
monopole mass by the ratio
\be
{{\mu L_i}\over {m_M}} \sim 10^{-2}
\ee
so the monopoles will move non-relativistically. 

In order to estimate the radiation rate we can assume that the
monopoles are moving in straight lines.  In fact, at the time of
string formation the monopoles will have thermal velocities, so that
in general the system will be formed with some non-zero angular
momentum.  However, in general this will be small compared to the
linear motion due to the string tension, so we will ignore it, except
to note that the monopoles will pass by without collision.  The half
oscillation of one monopole is parameterized by
\be
x(t) = (2 a L)^{1/2} t - {1\over 2} a t^2
\ee
with $a = \mu / m_M$ and $0 < t < (8L/a)^{1/2}$. Using the 
quadrupole approximation,\footnote{The fully relativistic situation
was considered in \cite{Martin:1997cp}.} the rate of energy loss of 
the system is
\be
{{dE}\over{dt}} = {288 \over 45} G \mu^2 \left({{\mu L}\over {m_M}}\right)
\ee
Since $\mu L$ is the energy in the string, we can integrate this 
equation to obtain 
\be
L = L_i e^{- t/\tau_g}
\ee
with
\be\label{eqn:tau}
\tau_g =  {45 \over 288} {m_M \over G \mu^2 } =
{45\over 288}{m_{pl}^2 m_M \over T_s^4}\,.
\ee
The lifetime of the state will thus be $\tau_g\ln (L_i/r_M)$, where
$r_M\sim m_M^{-1}$ is the radius of the monopole core.  For $T\sim
100 \, \text{GeV}$ and $m_M\sim 10^{14} \, \text{GeV}$, Eq.\ (\ref{eqn:tau}) gives
$\tau_g\sim 10^{17}\, \text{sec}$,
comparable with the age of the universe.

This suggests that the bound system formed by a monopole-antimonopole pair
connected by a string can slowly decay gravitationally, and release
 the energy stored in the monopole in a final annihilation
when the two monopole cores become close enough. 

\section{Conclusions}

We have shown that is not possible to construct a consistent model for 
the origin of the UHECR based on the electromagnetic decay and final 
annihilation of magnetic monopole-antimonopole bound states formed in the 
early universe. We have obtained an upper limit  for the monopolonium 
density today, taking into account its enhancement in the galactic halo 
and the maximum average free monopole density consistent with the Parker 
limit. Due to the small radiative capture cross section for the monopoles 
and the rapid expansion of the universe, the maximum density of 
monopolonium is many orders of magnitude below the  
concentration required to explain the highest energy cosmic ray events.

We then proposed a different scenario in which the monopoles are connected 
by strings that form at a relatively low energy. This mechanism solves 
the problem of the inefficiency of the binding process, since every
 monopole will be attached to an antimonopole at the other end of the
 string. Due to the confinement of the monopole flux inside of the 
string , the main 
source of energy lose for these bound systems will be gravitational radiation.
If we assume a monopole mass of  $10^{14}\, \text{GeV}$ and a string 
energy scale of the order of $100 \, \text{GeV}$, the lifetime of
 the bound states would be comparable with the age of the universe, 
making them a possible candidate for the origin of the 
ultra-high-energy cosmic rays.

\section{Acknowledgments}
We would like to thank Alex Vilenkin for suggesting this line of work,
and Xavier Siemens and Alex Vilenkin for helpful
conversations. This work was supported in part by funding provided by
the National Science Foundation.  J. J. B. P. is supported in part by the
Fundaci\'on Pedro Barrie de la Maza.

\end{document}